\documentclass{PoS}

\title{Higgs plus one jet production at NNLO}

\ShortTitle{Higgs plus one jet production at NNLO}

\author{Xuan Chen\\
 Peking University\\
 E-mail: \email{xuan.chen@pku.edu.cn}}

\author{Thomas Gehrmann\\
 University of Z\"urich\\
 E-mail: \email{thomas.gehrmann@uzh.ch}}

\author{Nigel Glover\\
 University of Durham\\
 E-mail: \email{e.w.n.glover@durham.ac.uk}}

\author{\speaker{Matthieu Jaquier}%
\\
University of Freiburg\\
E-mail: \email{matthieu.jaquier@uni-freiburg.de}}

\abstract{We present the computation of Higgs boson plus one jet production at NNLO at the LHC. The computation is carried out in the heavy-top-quark approximation and using the antenna subtraction formalism to handle the IR divergent structure of the individual contributions. All channels and all colours are included. We illustrate aspects of the subtraction formalism and present results for the total cross section as well as distributions in the kinematical variables using a realistic experimental setting. We make a comparison between our results and recent ATLAS data for $H+j$ production.}

\FullConference{12th International Symposium on Radiative Corrections (Radcor 2015) and LoopFest XIV (Radiative Corrections for the LHC and Future Colliders)\\
15-19 June 2015\\
UCLA Department of Physics \& Astronomy Los Angeles, CA, USA}

\begin{document}
\section{Introduction}
After the discovery of the Higgs boson at the CERN LHC \cite{ATLASHDisc, CMSHDisc}, attention has shifted towards precision studies of its properties. In the context of the search for physics beyond the standard model, the precise way the Higgs mechanism is realised might help to narrow down the list of possible candidate models. In particular, the precise measurement of spin, charge and parity of the Higgs boson as well as its couplings to the standard model particles provide valuable input for future searches.

One key property for characterising Higgs boson events is the number of jets produced in the event.
In gluon fusion, the dominant production mode at the LHC, the Higgs boson is frequently produced on its own, recoiling against 
soft radiation.  In some fraction of the events, this radiation leads to one or more hard jets. 
 The presence of these jets 
plays a key role in the background rejection for many Higgs boson studies. In particular one needs to understand the effect on the signal of $p_T$ cuts used to define the jets~\cite{ATLASHDisc, CMSHDisc, PTCut}.
In certain cases, like for instance $H\rightarrow WW$ processes, the events are classified into jet bins to allow for dedicated treatment of the different backgrounds \cite{ATLASHDisc, CMSHDisc, HWWlnulnu}. 
There, knowledge of the $H$ and $H+j$ cross sections at the same order of accuracy is important to take account of migration between the jet bins \cite{Binning} in a consistent way. The Higgs $p_{\mathrm{T}}$ spectrum is also a central quantity when studying the decay of a boosted Higgs boson into a fat jet, as it is done when tagging $H\rightarrow b\bar{b}$ decays \cite{JetSubStruc}.

In the standard model the Higgs boson does not couple directly to gluons, and so the interaction is mediated at the Born level through quark loops, where the dominating contribution comes from the top quark \cite{higgstop}. In the limit of a very heavy top quark, the loop can be integrated out and one obtains a point-like interaction between the Higgs boson and gluons \cite{effth}. This approximation reduces the number of loops in a calculation by one and is valid as long as the scales in the process do not exceed $2m_{\mathrm{t}}$ \cite{ptdist}. Using this effective theory, the $H+j$ process has been computed at next-to-leading order (NLO) some time ago \cite{HJNLO} and has since then also been augmented with finite $m_{\mathrm{t}}$ effects \cite{HJNLOMTfin}. More recently, two independent calculations of the same process have been performed at NNLO \cite{HSector, HNJettiness}. A computation of the fiducial cross section as well as kinematical distributions has also been performed for $pp\rightarrow H+j\rightarrow \gamma\gamma+j$ using ATLAS cuts at 8 TeV as well as for $pp\rightarrow H+j\rightarrow W^+W^-+j$ in the CMS setup at 13 TeV \cite{HFiducial}. We present here a new calculation of the complete NNLO corrections to $H+j$ production, extending our earlier results \cite{ourpaper} at this order for the gluons-only subprocess.

\begin{figure}[t]
 \begin{center}
  (a)\includegraphics[trim= 20 60 0 200, clip, width=2.5cm]{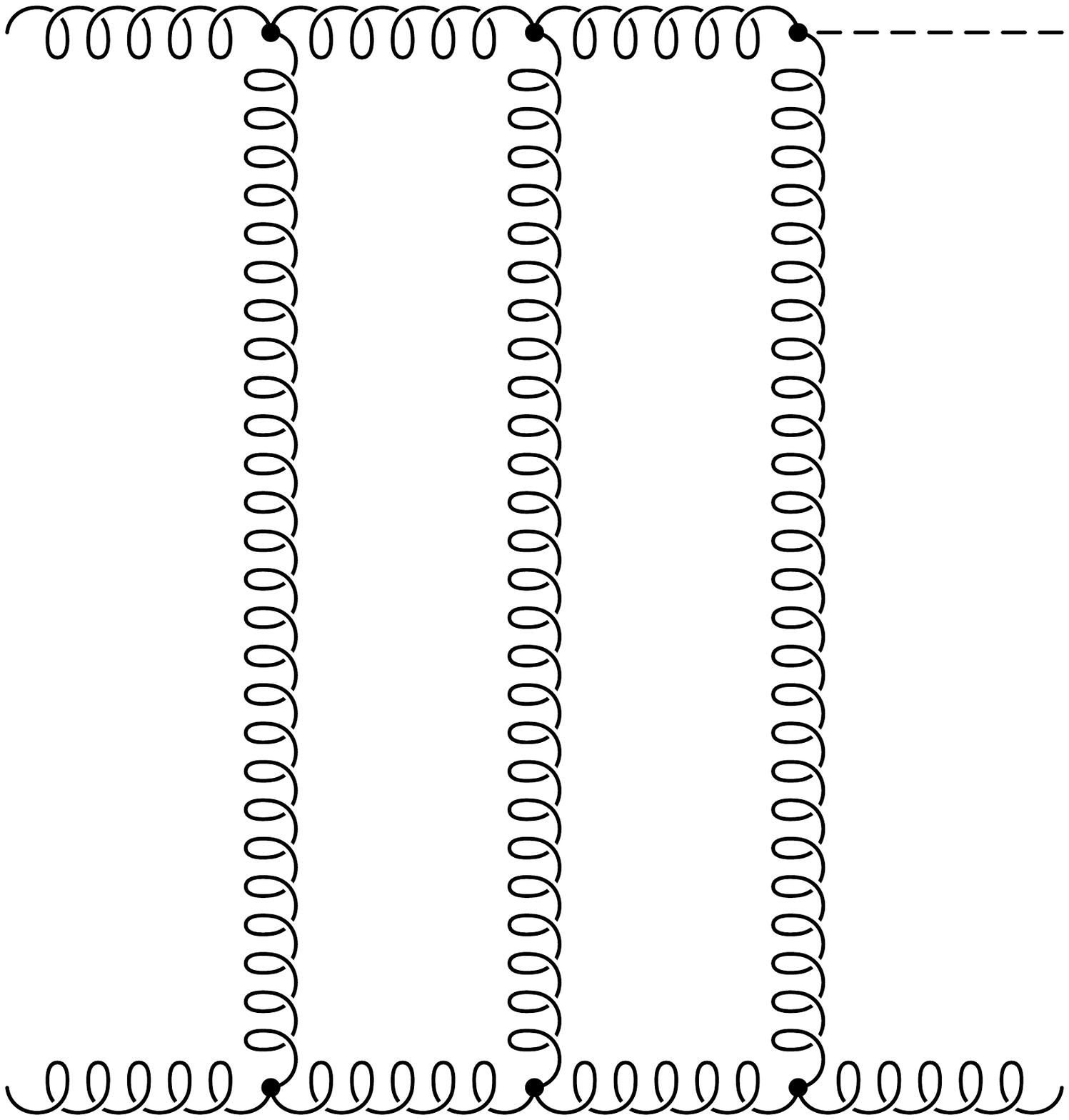}$\qquad$
  (b)\includegraphics[trim= 20 60 0 200, clip, width=2.5cm]{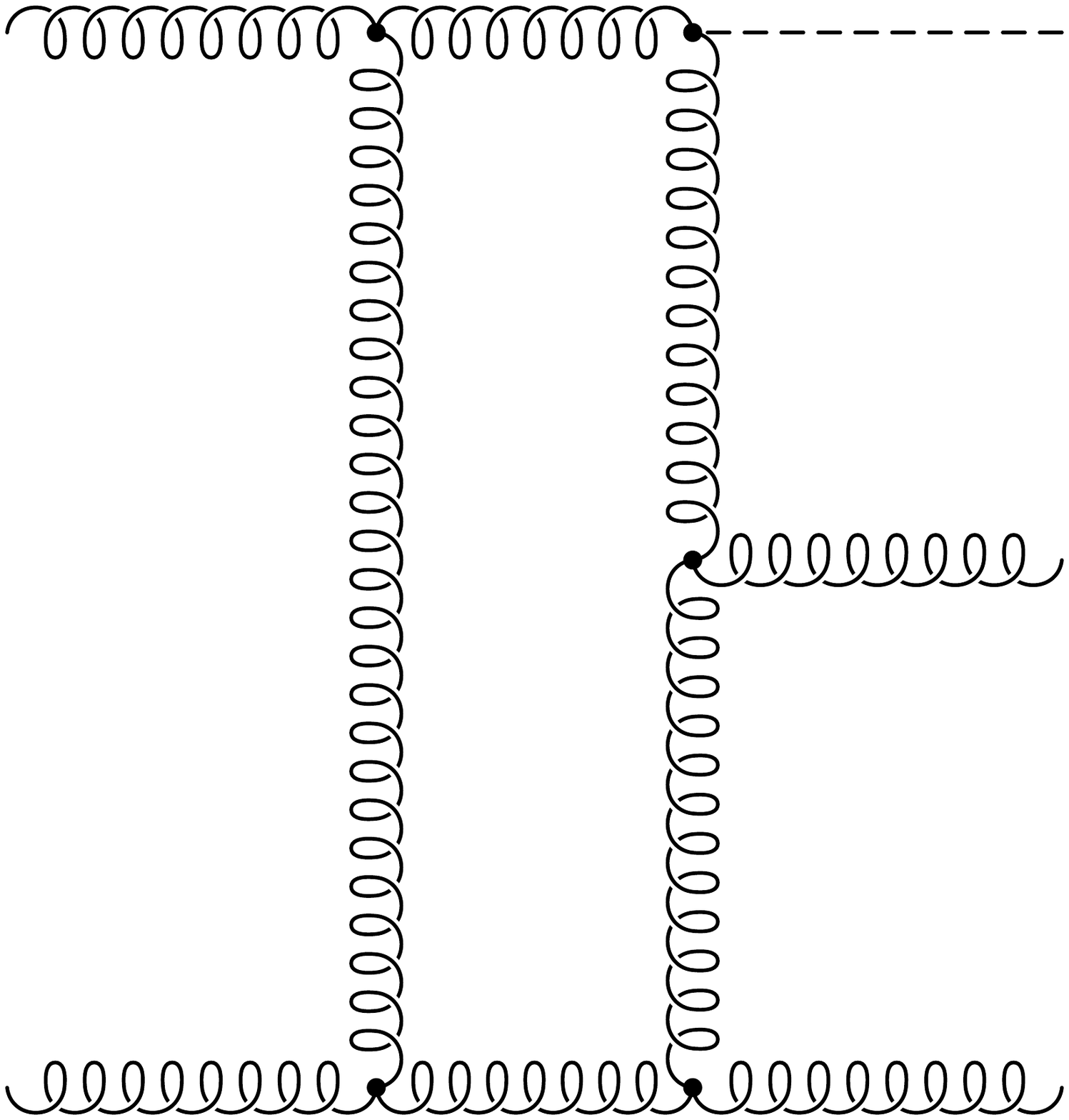}$\qquad$
  (c)\includegraphics[trim= 20 60 0 200, clip, width=2.5cm]{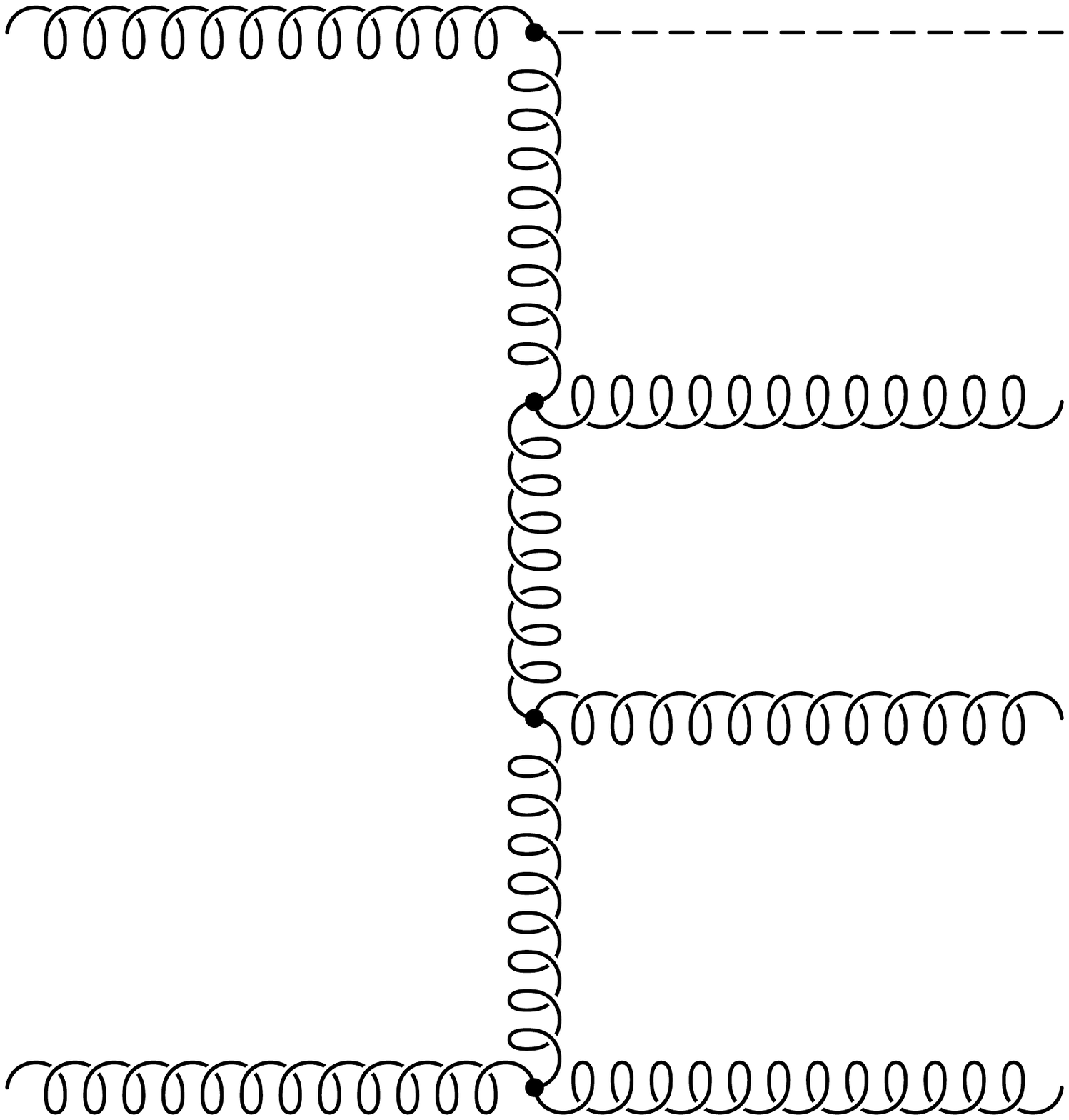}
  \caption{Representative Feynman diagrams in the effective theory for                                                                                                                                               
  (a) two-loop Higgs-plus-three-gluon amplitudes (b)                                                                                                                                                                 
  one-loop Higgs-plus-four-gluon amplitudes and (c)                                                                                                                                                                  
  tree-level Higgs-plus-five-gluon amplitudes.\label{fig:FD}}
 \end{center}
\end{figure}

 A major difficulty arising in higher-order calculations is the occurence of several contributions which are separately ill-defined, and only lead to a sensible result once added together in a consistent way. At NNLO, the cross section receives contributions from processes which, with respect to the leading order, feature two additional external legs (double real emission, \cite{rrh}), one external leg and one internal loop (real virtual, \cite{rvh}) or two additional internal loops (double virtual, \cite{vvh}), see Figure~\ref{fig:FD}. Each of these pieces suffers from infrared (IR) divergences: in virtual parts, the integral over the loop momentum includes configurations where it is unresolved (soft or collinear) with respect to one of the external momenta. The resulting divergences are handled using dimensional regularisation and expressed in terms of poles in $D-4$. The subsequent integration over the appropriate phase space of the various contributions is performed numerically due to the presence of a jet function. There, unresolved external momenta lead to further IR divergences which need to be removed to ensure a well defined result. The extraction of IR divergences and their cancellation by virtue of the KLN theorem \cite{KLN} can be achieved in a variety of ways. The two previous $H+j$ calculations have employed sector improved residue subtraction \cite{Stripper} and the N-Jettiness slicing method \cite{NJettiness}. The N-Jettiness method was also used recently in the computation of $W+j$ \cite{WNJettiness} and $Z+j$ \cite{ZNJettiness} at NNLO accuracy. Further subtraction methods are $Q_\mathrm{T}$-subtraction \cite{qtsub}, sector decomposition \cite{secdec} and antenna subtraction \cite{antsub}. In this contribution, we report on the computation of $H+j$ at NNLO using the antenna subtraction formalism. The antenna subtraction formalism has been employed in the computation of the NNLO corrections to $e^+e^-$-production \cite{epemGGGH,epemWZ}, to the production of heavy particles \cite{heavyPart}, to $Z+j$ \cite{ZJAnt} and to several leading colour contributions to dijet production \cite{Dijets}. This computation will allow for a benchmarking of the different subtraction schemes and lead to valuable insight on the difficulties arising in NNLO computations. After a brief review of the antenna subtraction formalism in section 2, we will present its implementation in section 3 together with results for the total cross section and differential distributions before concluding in section 4.

\section{Antenna subtraction}
It is well-known that in unresolved momentum configurations, a QCD amplitude factorises into products of universal unresolved factors which contain the singularities and reduced amplitudes which depend only on the resolved kinematics \cite{soft, split}. At NLO in collinear limits, the divergences are described by splitting functions, whereas divergences in soft limits are contained in eikonal factors. This continues at NNLO and holds also for the full matrix element. The aim of a subtraction formalism like antenna subtraction is to reproduce this behaviour for the real radiation contributions in unresolved kinematics. It is encapsulated in subtraction terms which, when added to the corresponding matrix elements, make them well-defined and integrable over the whole phase space, thus allowing for numerical integration. The subtraction terms can then be integrated analytically over the phase spaces of the unresolved partons. The integrated subtraction terms obtained in this way have their singularities expressed in terms of poles in $D-4$, such that they can be combined with the virtual contributions to subtract the pole structure of the loop matrix elements.

The central object in the antenna subtraction scheme is the $m$-parton $l$-loop antenna function $X_m^l(p_a,p_2,\ldots,p_{m-1},p_b)$. This function features the full IR divergent behaviour of the partons $p_2,\ldots,p_{m-1}$ in the colour ordering $\{p_a,p_2,\ldots,p_{m-1},p_b\}$, where $p_a$ and $p_b$ are labeled as the hard radiators. At NLO the only unresolved singularities one has to deal with are single unresolved limits of tree matrix elements, such that one has $m=3$ and $l=0$. At NNLO one can furthermore have $m=4$ and $l=0$ to describe the double unresolved singularities of tree matrix elements as well as $m=3$ and $l=1$ to take account of single unresolved limits of one loop matrix elements that appear in the real-virtual contribution. Following the factorisation behaviour of QCD amplitudes, a typical subtraction term then has the form
\begin{equation}
 X_m^l(p_a,p_2,\ldots,p_{m-1},p_b)|M_n(\ldots,\tilde{p}_a,\tilde{p}_b,\ldots)|^2J_n(\ldots,\tilde{p}_a,\tilde{p}_b,\ldots),
\end{equation}
where the reduced matrix element $M_n$ (possibly with loops) and the jet $J_n$ function depend only on the resolved momenta $\tilde{p}_a,\tilde{p}_b$. These are given through mappings $\{p_a,p_2,\ldots,p_{m-1},p_b\}\rightarrow\{\tilde{p}_a,\tilde{p}_b\}$ which interpolate between all unresolved configurations of the partons $p_2,\ldots,p_{m-1}$ with the hard radiators \cite{mappings}.

One can then perform the integral over the phase space of the unresolved momenta to obain the integrated antenna $\mathcal{X}_m^l(p_A,p_B)$, which feature explicit poles up to order $2(m-2+l)$ in $D-4$. These integrated antennae can be related to the infrared singularity operator $\mathbf{I}_l(p_A,p_B)$ introduced by Catani to describe the singularity structure of partons emitted between the dipole formed by $p_A$ and $p_B$ \cite{catani}. It can be shown that by combining integrated antennae with an adequate subset of the mass factorisation counterterms, which originate from the renormalisation of the PDFs, one can define operators $\mathbf{J}_l$ such that 
\begin{equation}
 \mathbf{J}_l(p_A,p_B) = \mathbf{I}_l(p_A,p_B) + Finite.
\end{equation}
Using that correspondence, one can always rearrange the integrated antennae together with the mass factorisation counterterms into a structure which closely matches the pole structure of the virtual matrix elements as predicted by Catani, allowing for a controlled cancellation \cite{antsub}.

It is interesting to note that once this has been done for an appropriate set of processes, one obtains a dictionary between the unresolved singularities of real matrix elements and the poles of the virtual ones. The knowledge of the pole structure at one loop can then be used to provide a guideline for the construction of the subtraction term for the real contribution. This perspective still holds true at NNLO.
\section{Implementation and results}
We have implemented all relevant subprocesses for $H+j$ at NNLO in a parton-level event generator, relying on the antenna subtraction method to cancel the implicit and explicit IR divergences appearing in the matrix elements of the various contributions. This program allows for the computation of all IR-safe observables related to $H+j$ final states to NNLO accuracy. Renormalisation and factorisation scale can be chosen on an event-by-event basis. The decay of the Higgs boson to two daughter particles is included, such that we can make predictions for realistic experimental data by providing the appropriate event selection criteria and Higgs decay matrix elements. As an application, we study $H\rightarrow\gamma\gamma$ production together with one jet at NNLO accuracy at an 8 TeV LHC. Due to the small width of the Higgs boson the decay to two photons factorises, and can therefore be implemented by multiplying the event weight by the corresponding branching ratio. We provide precise differential distributions in various kinematical variables and compare them with data recently published by the ATLAS collaboration \cite{ATLASgamgam}.

For our computation we use NNPDF2.3 parton distribution functions. We set $m_H=125$ GeV for the Higgs boson mass and $\mu_R=\mu_F=m_H$ for the renormalisation and factorisation scales. The theory error is determined by simultaneously varying $\mu_R$ and $\mu_F$ to $2m_H$ and $\frac{m_H}{2}$. Following ref. \cite{ATLASgamgam}, we cluster jets with the anti-$k_T$ algorithm with $R=0.4$. The selection criteria are $p_{T,J}>30$ GeV and $|\eta_J|<4.4$ for the transverse momentum and rapidity of the jets, respectively. Photons must satisfy $|\eta_{\gamma}|<2.37$ and $p_{T,\gamma_1}>43.75$ GeV and $p_{T,\gamma_2}>31.25$ GeV for the leading and subleading photon transverse momenta respectively. We furthermore require a minimum distance of 0.4 between jets and photons on the $\eta$-$\phi$ plane. With this, we obtain for the total cross section
\begin{equation}
 \sigma_{NNLO}=9.40^{+0.65}_{-0.89}\ \mathrm{fb},
\end{equation}
where the upper (lower) value corresponds to the scale choice $\mu_R=\mu_F=\frac{m_H}{2}$ ($\mu_R=\mu_F=2m_H$). We observe the NNLO corrections reduce the scale uncertainty to around 10\%. This fixed order perturbative result is substantially lower than the measured experimental value quoted in \cite{ATLASgamgam}, of $21.5\pm5.3(\mathrm{stat})^{+2.4}_{-2.2}(\mathrm{syst.})\pm0.6(\mathrm{lumi})$ although the experimental uncertainties are large. In order to check our result, we compare it with \cite{HFiducial}, by removing contributions where both initial partons are either a quark or an antiquark from the NNLO correction. We obtain
\begin{equation}
 \sigma'_{NNLO}=9.44^{+0.59}_{-0.85}\ \mathrm{fb},
\end{equation}
in good agreement with \cite{HFiducial}.

We compute distributions in the kinematical variables and compare them with 8 TeV data from ATLAS \cite{ATLASgamgam}.
\begin{figure}[t!]
 (a)$\qquad\qquad\qquad\qquad\qquad\qquad\qquad\qquad\qquad$(b)\newline
 (c)
 \includegraphics[width=0.45\textwidth]{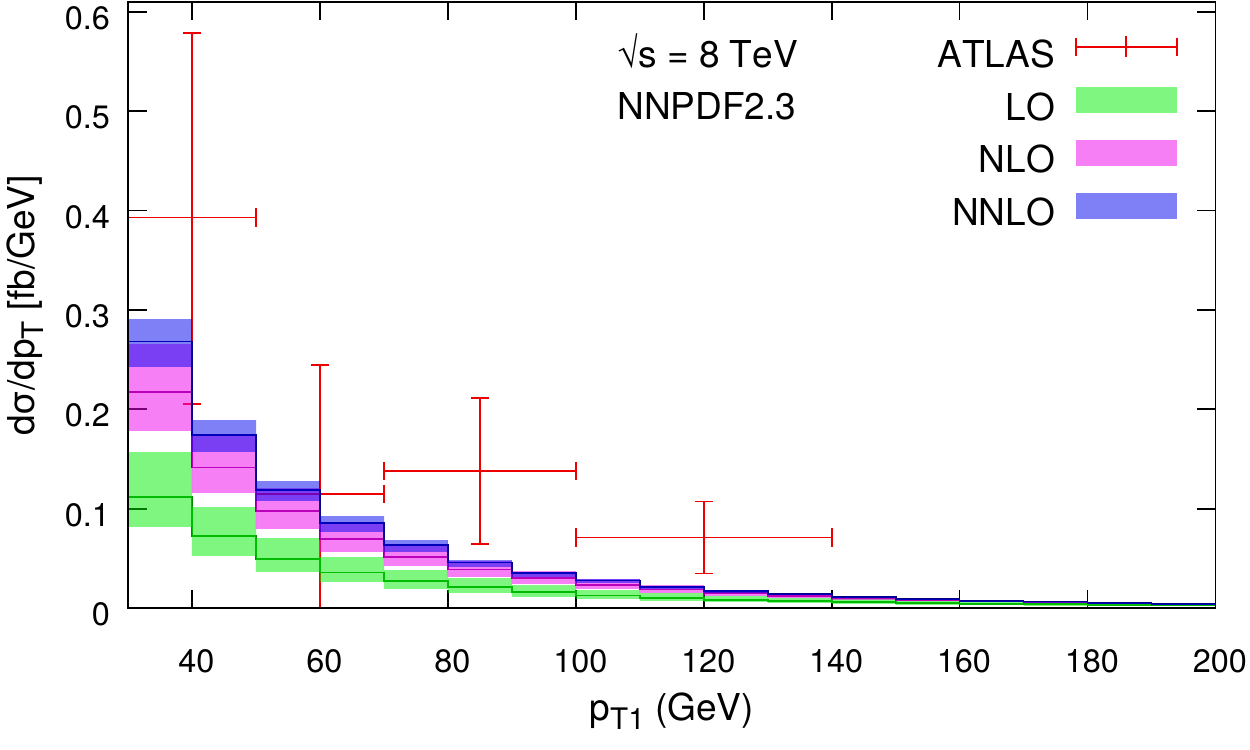}
 (d)
 \includegraphics[width=0.45\textwidth]{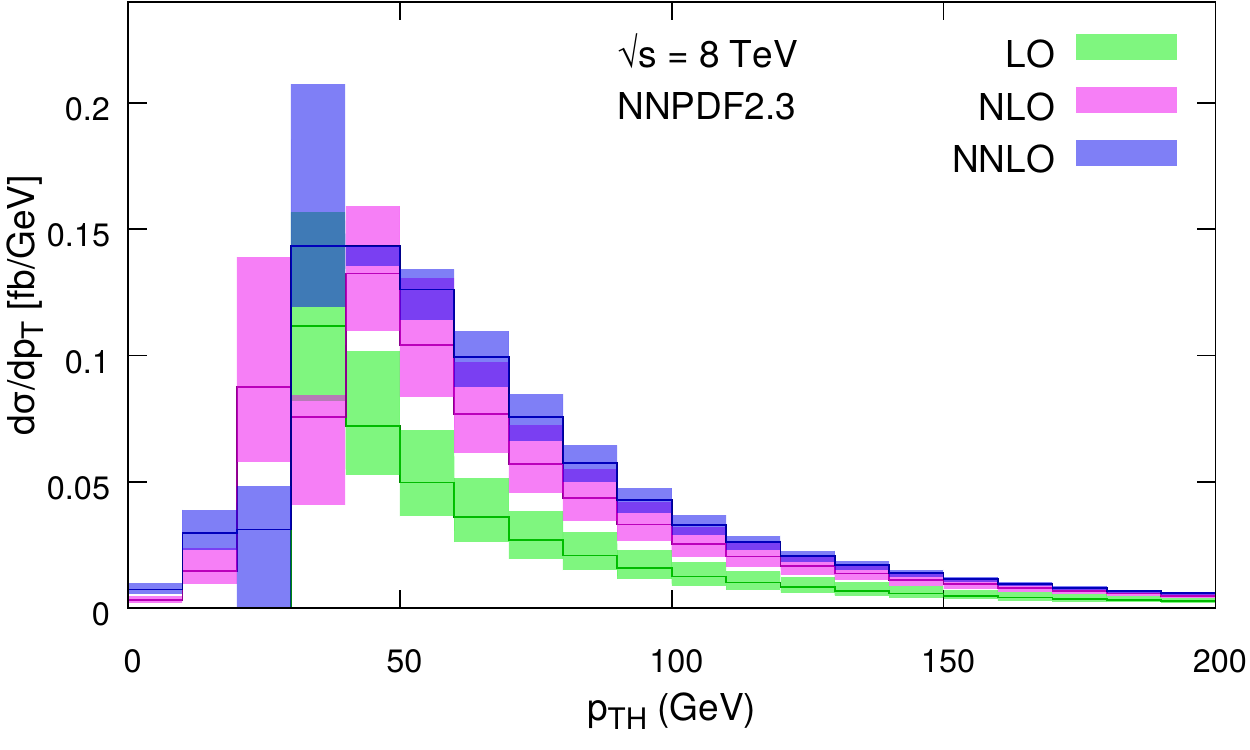}\vspace{3pt}\newline
 \phantom{(c)}
 \includegraphics[width=0.45\textwidth]{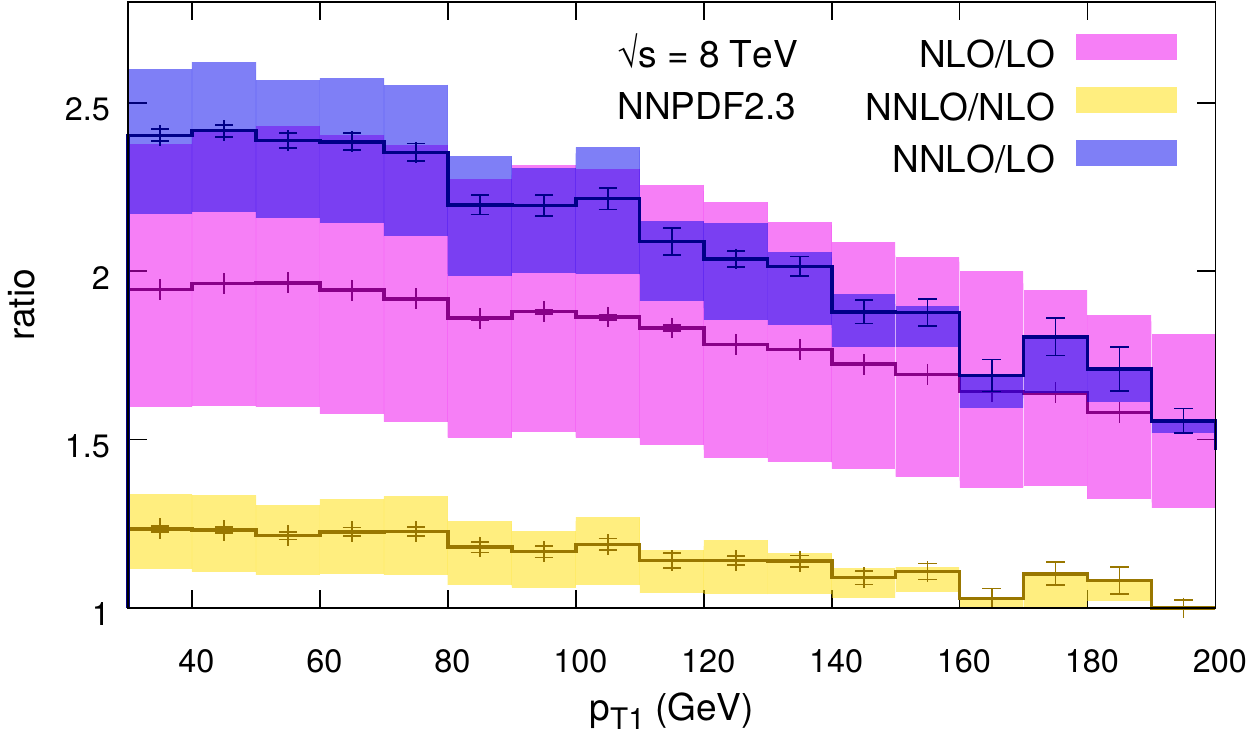}
 \phantom{(d)}
 \includegraphics[width=0.45\textwidth]{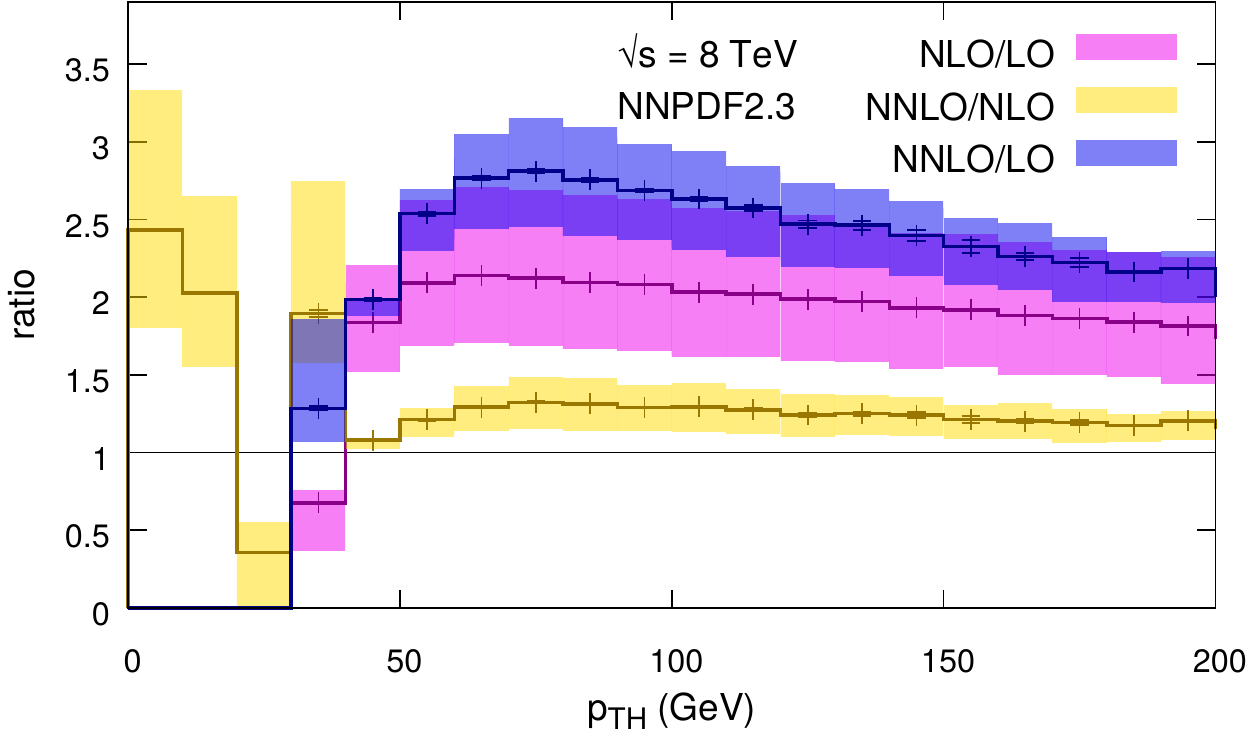}\vspace{-4pt}
 \caption{(a) Transverse momentum distribution of the leading jet and (b) the Higgs boson in inclusive $H+j\rightarrow\gamma\gamma+j$ production in $pp$ collisions at $\sqrt{s}=8$ TeV at LO, NLO and NNLO as computed in this paper and measured by the ATLAS experiment and (c), (d) ratios of different perturbative orders, NLO/LO, NNLO/LO and NNLO/NLO.}\label{fig:pt}\vspace{5pt}
 (a)$\qquad\qquad\qquad\qquad\qquad\qquad\qquad\qquad\qquad$(b)\newline
 \phantom{(a)}
 \includegraphics[width=0.45\textwidth]{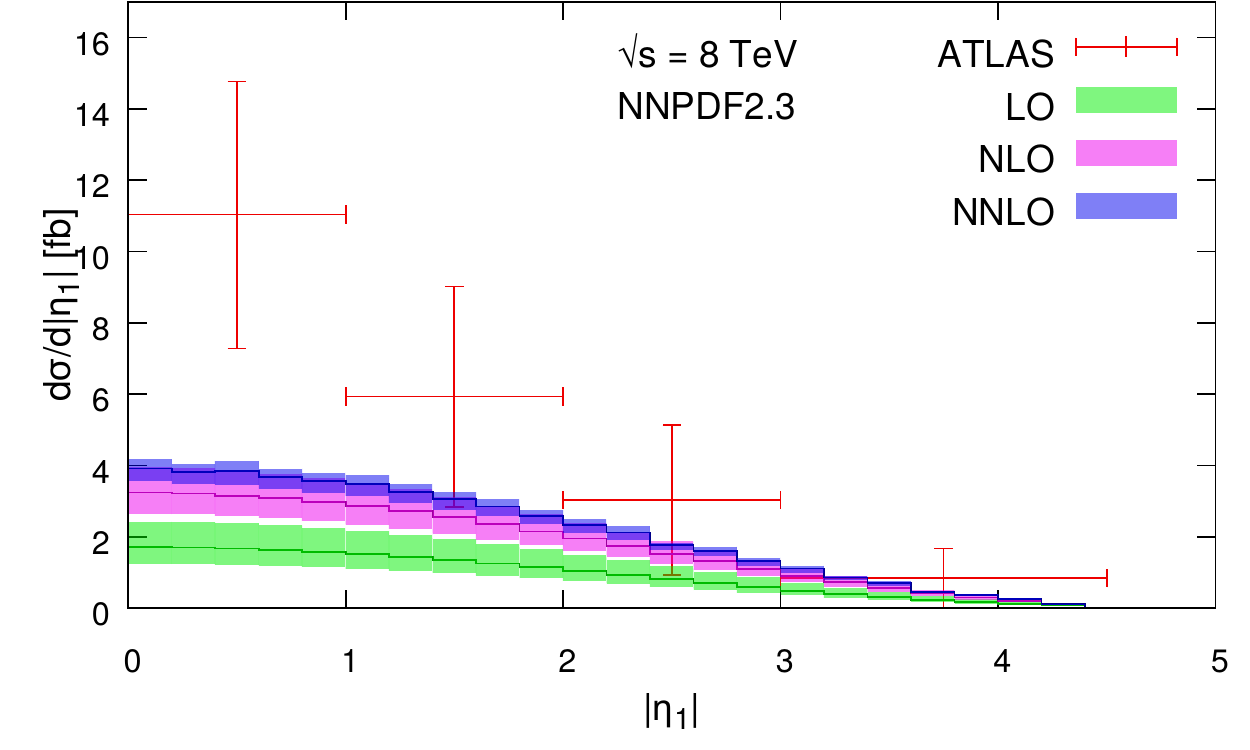}
 \phantom{(b)}
 \includegraphics[width=0.45\textwidth]{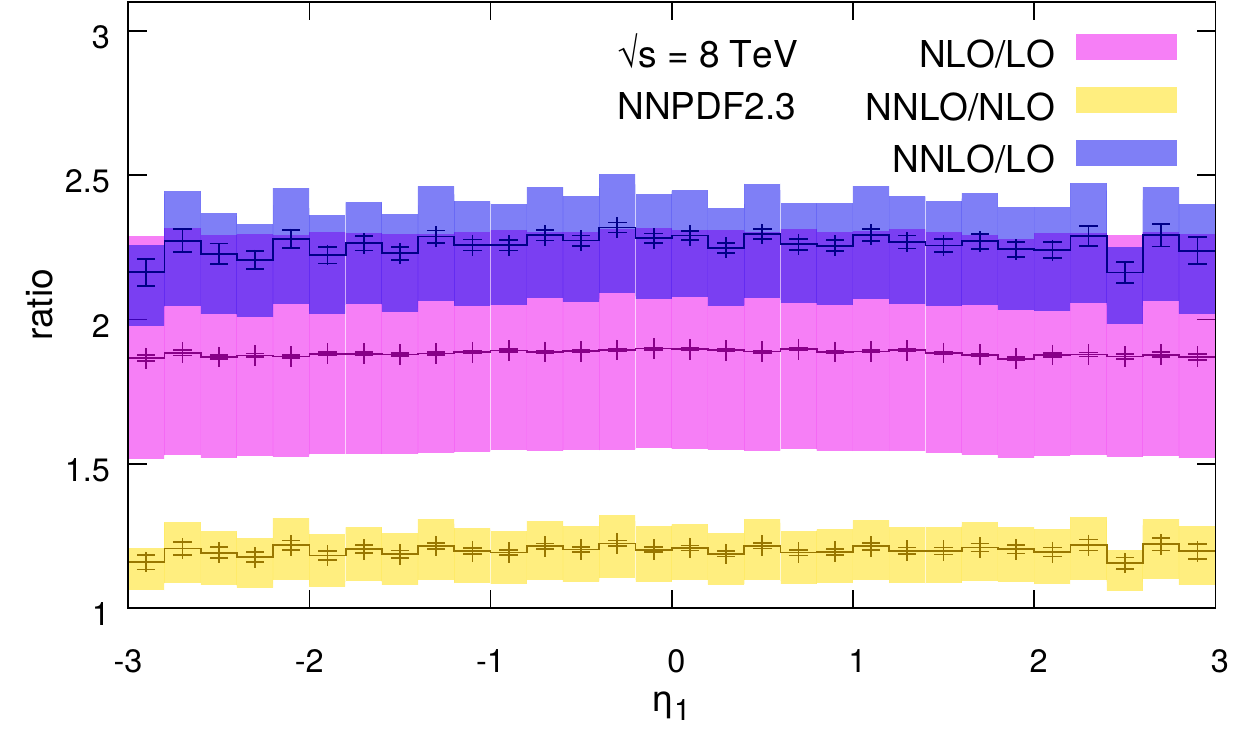}
 \caption{(a) Rapidity distribution of the leading jet in inclusive $H+j\rightarrow\gamma\gamma+j$ production in $pp$ collisions at $\sqrt{s}=8$ TeV at LO, NLO and NNLO as computed in this paper and measured by the ATLAS experiment and (b) ratios of different perturbative orders, NLO/LO, NNLO/LO and NNLO/NLO.}\label{fig:eta}
\end{figure}
 In fig. \ref{fig:pt} (a) and (b) we display transverse momentum distributions for the leading jet and for the Higgs boson, where we evaluate the error band as the scale variation envelope using the same scale settings as for the total cross section. The LO and NLO distributions are evaluated using NNLO PDFs to include some higher order effects and to better access the impact of the genuine NNLO correction. We also show in fig. \ref{fig:pt} (c) and (d) ratio plots between the different perturbative orders, where in the error bands the denominator is always evaluated at the central scale, such that the band reflects the scale variation of the numerator only. The error bars on the ratio plots indicate the numerical integration error for the corresponding bin. We observe sizeable NNLO corrections to the distributions, while their shape is only slightly modified. For the leading jet, the change is most pronounced at low $p_\mathrm{T}$ values, where the distributions become enhanced by a factor of $\sim$1.2, while for higher $p_\mathrm{T}$ the differential K-factor approaches 1. There however, the top quark loop mediating the interaction between the Higgs boson and gluons starts being resolved. The effective theory therefore breaks down, and top quark mass effects would become important. For the Higgs boson we observe at low $p_\mathrm{T}$ the typical fluctuations which take their origin in the fact that at LO the Higgs is kinematically bounded due to the $p_\mathrm{T}$ cut on the single jet in the events - the well known Sudakov shoulder. At intermediate $p_\mathrm{T}$ we observe an enhancement with a factor of up to $\sim$1.3, which decreases to $\sim$1.2 for high $p_\mathrm{T}$. The remaining scale uncertainty is between 6-12\% for the leading jet $p_\mathrm{T}$, and 10-15\% for the Higgs boson $p_\mathrm{T}$ using scale variation as described above. Comparing the NNLO predictions to the ATLAS data, it appears that for the leading jet $p_\mathrm{T}$ the NNLO prediction lies within the experimental error at lower $p_\mathrm{T}$ values, while underestimating the experimental values at higher $p_\mathrm{T}$. This is also even more the case for the Higgs $p_\mathrm{T}$ distribution, which is noticably lower than the ATLAS data points.  We expect that the apparent tension between theory and experiment will be resolved once more data is collected in Run 2.
 
A similar behaviour can be observed in fig. \ref{fig:eta}, where the distribution in rapidity of the leading jet is displayed. The NNLO corrections enhance the distribution by a quasi constant factor of $\sim$1.2. The residual NNLO theory uncertainty is about 11\% over the whole rapidity range. At high rapidity values, the prediction agrees rather well with the experimental data, while being significantly lower at central rapidities. 

\section{Conclusions}
In this contribution, we reported on the complete computation of inclusive $H+j$ production at NNLO at an 8 TeV LHC. The calculation is performed using the heavy-top-quark approximation to reduce the number of loops by one and relies on the antenna subtraction formalism to control the various infrared divergences appearing in the different contributions at this order in perturbation theory. It is implemented in a Monte Carlo event generator which allows for the computation of any user defined observable in $H+j$ production.

We implemented the decay of the Higgs boson to two photons and calculated the total cross section as well as various differential distributions in the setup of the ATLAS experiment. We find the total cross section agrees well with the results of Caola, Melnikov and Schulze~\cite{HFiducial} for the same setup. Both the total cross section and the distributions have substantial NNLO corrections, while the scale uncertainty for these observables is significantly reduced, leaving a residual scale uncertainty of $\sim$10\% for the total cross section and up to $\sim$15\% in the distributions. The change in the shape of the distributions due to the NNLO correction is quite modest and mostly amounts to an overall normalisation factor. We compare the results from our calculation with data from ATLAS and find them to fall below the (statistically limited) experimental values, as well for the total cross section as for the differential distributions.

Finally, we note that our implementation is fully exclusive for the $H+j$ final state and can be easily upgraded for other Higgs decay processes. It will therefore provide an important and flexible tool for precision studies of the Higgs boson in the future high-luminosty run of the LHC.

\end{document}